# The AI generation gap: Are Gen Z students more interested in adopting generative AI such as ChatGPT in teaching and learning than their Gen X and Millennial Generation teachers?

Authors: Cecilia Ka Yuk Chan[1]* & Katherine K. W. Lee[1]

Affiliation[1]: The University of Hong Kong

Address: Centre for the Enhancement of Teaching and Learning (CETL), Room CPD-1.81, Centennial Campus, The University of Hong Kong, Pokfulam, Hong Kong

* Corresponding author. Email: cecilia.chan@cetl.hku.hk

Email: kathkw@connect.hku.hk

**Abstract**

This study aimed to explore the experiences, perceptions, knowledge, concerns, and intentions of Gen Z students with Gen X and Gen Y teachers regarding the use of generative AI (GenAI) in higher education. A sample of students and teachers were recruited to investigate the above using a survey consisting of both open and closed questions. The findings showed that Gen Z participants were generally optimistic about the potential benefits of GenAI, including enhanced productivity, efficiency, and personalized learning, and expressed intentions to use GenAI for various educational purposes. Gen X and Gen Y teachers acknowledged the potential benefits of GenAI but expressed heightened concerns about overreliance, ethical and pedagogical implications, emphasizing the need for proper guidelines and policies to ensure responsible use of the technology. The study highlighted the importance of combining technology with traditional teaching methods to provide a more effective learning experience. Implications of the findings include the need to develop evidence-based guidelines and policies for GenAI integration, foster critical thinking and digital literacy skills among students, and promote responsible use of GenAI technologies in higher education.

## 1. Introduction

Generation Z (Gen Z) students have largely replaced Millennials in undergraduate programmes, with institutions of higher education now primarily enrolling students from the former (Seemiller & Grace, 2016; Shatto & Erwin, 2016). With educators welcoming a new cohort of students to campus, there is a growing concern regarding how to effectively teach this 'always-on' generation; for example, a study by Pearson (2018) showed that almost half of all Gen Z-ers (47%) spend a minimum of three hours daily on YouTube.

The Gen Z population, much like its predecessors – the Silent and Baby Boomer generations, followed by Generation X (Gen X) and Generation Y (also known as Millennials) – has its own unique, distinct characteristics that have been shaped by information communication technologies, social and cultural shifts, and financial volatility. As such, it is crucial for higher education institutions to effectively engage with Gen Z, in order for scholars, teachers, and university staff to understand their aforementioned characteristics (Seemiller & Grace, 2017; Shatto & Erwin, 2016; Shorey et al., 2021) and in turn, effectively and ethically integrate generative AI (GenAI) technologies into the curriculum. The changing student population and simultaneous technological advances, including GenAI, should be a stimulus for evaluation and potential modification of policy and pedagogical approaches within the traditional classroom and experiential learning settings. Furthermore, it is imperative to offer support to Gen X and Millennial teachers on GenAI technologies, by examining their perceptions and concerns, in order to reduce the gap of their expectations for promoting seamless integration and collaboration, ultimately improving the overall learning experience and harnessing the full potential of AI-driven educational tools. Thus, the research questions for this study are:

1. Regarding the use of GenAI in higher education, what are the current experiences, perceptions, knowledge, and concerns of Gen Z students and of Gen X and Gen Y teachers?
2. Regarding the use of GenAI in higher education, how do Gen Z students compare to Gen X and Gen Y teachers in terms of their current experiences, perceptions, knowledge, and concerns?
3. What are the current intentions and concerns of Gen Z students and of Gen X and Gen Y teachers, along with any accompanying reasons, towards using GenAI in higher education?

## 2. Literature Review

### 2.1. Generative AI Technologies in Higher Education

GenAI technologies, particularly large-scale language models like ChatGPT, have the potential to reshape higher education by offering new possibilities for enhancing teaching, learning, and student engagement (Hu, 2023). These technologies leverage advanced natural language processing and deep learning techniques to generate human-like text based on input data, enabling them to engage in meaningful conversations and provide relevant, context-aware information (Goodfellow et al., 2014).

As such, the development of AI-driven virtual tutors that provide students with instant, personalized guidance and feedback on various subjects (Alam, 2022; Celik et al., 2022; Terzopoulos & Satratzemi, 2019) is a significant innovation for education. This on-demand assistance can help address individual learning gaps, reinforce understanding, and support self-paced learning, particularly in remote and online learning environments, allowing anytime, anywhere for anyone. GenAI can also facilitate peer collaboration by connecting students with diverse backgrounds, interests, and expertise. For example, it can be integrated into communication platforms, such as forums and messaging apps, to encourage knowledge sharing, group problem-solving, and interdisciplinary collaboration (McLaren, Scheuer, & Mikšátko, 2010; Sharma et al., 2023). This enhanced connectivity can contribute to the development of a more inclusive and dynamic learning community.

Another promising application of GenAI in higher education is its potential to assist educators in generating course materials, such as summaries, quizzes, and discussion prompts. By synthesizing and rephrasing existing content, it can save time and effort for educators, allowing them to focus on more complex aspects of course design and pedagogy. Based on a review by Chen et al., (2023) on AI in education, using a qualitative research approach, it was found that AI systems have enabled the customization and personalization of curriculum and content according to students' needs, leading to improved learning experiences, and overall educational quality.

Apart from pedagogical assistance, GenAI is also adept at handling administrative tasks. A recent study by Kumar & Raman (2022) surveyed 682 Indian business management students to assess their opinions on AI usage in higher education. Students had positive perceptions of AI, especially in administrative and admission processes. However, they were more hesitant about AI replacing faculty in teaching-learning processes. The study revealed that students' prior exposure to AI influenced their perceptions. Moreover, AI can support higher education institutions in creating a more personalized and engaging learning experience by analysing vast amounts of data to identify trends, patterns, and insights. These insights can be used to inform the development of targeted interventions, course materials, and learning strategies that cater to the diverse needs and preferences of students (Daniel, 2015).

Despite the many potential applications of GenAI in higher education, several challenges and considerations need to be addressed to ensure its responsible and effective integration. These include data privacy and security concerns, potential biases in AI algorithms, and the importance of human oversight in AI-driven decision-making processes

(Bisdas et al., 2021; Chan & Hu, 2023; Chan & Zhou, 2023; Chen et al., 2023; Essel et al., 2022; Gillissen et al., 2022; Jha et al., 2022). By acknowledging and addressing these concerns, higher education institutions can better harness the power of GenAI technologies to foster innovation and improve learning outcomes.

**2.2. Literature on the Generations**

A literature review comparing Baby Boomers, Gen X, Gen Y (Millennials), and Gen Z reveals significant differences in characteristics between these populations across various aspects such as teaching preferences, learning styles, technology usage, and communication methods. These generational differences can be attributed to distinct historical events, financial instability, social experiences, and technological advancements that have shaped each generation's upbringing and development (Hernandez-de-Menendez et al., 2020; Puiu, 2017; Wiedmer, 2015). Table 1 provides an overview of distinct generational traits and characteristics based on literature reviewed.

**2.2.1. Baby Boomers**

Baby Boomers, born between 1946 and 1960, grew up during a period of economic prosperity and optimism following World War II. As a result, they generally exhibit a preference for traditional lecture formats in educational settings and are characterized by teacher-centred learning, note-taking, and memorization (Zemke et al., 2000). Baby Boomers are also known for their willingness to share knowledge, patience, respect for traditions, and commitment to hard work. They value job security and organizational loyalty, with careers often defined by employers (Oblinger & Oblinger, 2005).

**2.2.2. Gen X**

In contrast, Gen X, born between 1960 and 1980, was raised during a time of economic uncertainty, marked by a shift towards individualism and self-reliance (Wiedmer, 2015). This generation tends to prefer a combination of traditional and technology-based methods for teaching and learning, and is characterized by collaborative, project-based, and real-world application approaches (Shamma, 2011). Gen X individuals are known to be more adaptable to change, open to diversity, and loyal to their profession rather than a specific employer. They value work-life balance and are considered "digital immigrants," who confidently use technology despite not growing up with it (Zemke et al., 2000).

**2.2.3. Gen Y (Millennials)**

Gen Y, also known as Millennials, born between 1980 and 1995, grew up during the rapid expansion of the internet and digital technology. Consequently, they have a natural affinity for technology and are often referred to as "digital natives" (Bencsik et al., 2016; Issac et al.,

2020). Millennials prefer interactive, self-paced, technology-based methods in education, and their learning style is characterized by collaboration and networking. They have high expectations for feedback and demand flexibility, mobility, and creativity in the workplace. This generation is more likely to work "with" organizations, rather than "for" them, and is known for their entrepreneurial spirit (Wiedmer, 2015).

### 2.2.4. Gen Z

As the majority of Baby Boomers have now entered retirement, there is a resulting predominance of experienced Gen X educators and junior Gen Y teachers in higher education institutions. Consequently, the focus of this study will be on the interactions between these two generational cohorts and the predominantly Gen Z undergraduate student population. Furthermore, as students are the primary recipients of education, a detailed exploration of their learning traits is warranted and thus presented below. This examination enables a better understanding of the unique characteristics and preferences of Gen Z learners, as well as the potential implications of intergenerational dynamics between Gen X and Gen Y educators and their students with the integration of GenAI technologies.

Gen Z, born between 1995 and 2012, is the first generation to grow up with constant access to digital technology and social media, resulting in a digital-first and technoholic mindset (Puiu, 2017). Seemiller and Grace (2016) emphasize that Gen Z has grown up in an era of rapid technological advancements, with digital devices and the internet being an integral part of their lives. This ubiquity of technology has shaped their learning preferences, communication styles, and expectations regarding education, as well as expectations for educational institutions to provide up-to-date technological resources and support.

The constant access to digital environments is a significant aspect that should not be underestimated (Turner, 2015) - this connection leads to distinctive learning attributes that set them apart from earlier generations (Seemiller & Grace, 2017; Shatto & Erwin, 2016). Gen Z tends to prefer hybrid learning approaches that incorporate technology and multimedia content, learning through images, videos, and audio instead of text (Seemiller & Grace, 2016; 2017), and studies such as those by Mosca, Curtis, and Savoth (2019) and Granitz, Kohli, and Lancellotti (2017) support the idea that Gen Z students prefer video-based learning over traditional textbook-based learning. Mosca, Curtis, and Savoth's survey (2019) in particular found that 98.5% of 133 university students agreed that videos help to bring in real-world situations; similarly, Granitz, Kohli, and Lancellotti's case study (2017) found that Gen Z students prefer video-based learning materials over traditional textbooks and found the latter more engaging.

Due to their immersion in digital communication platforms, Gen Z students are accustomed to instant and constant connectivity. They tend to favour short, concise messages and visual content over long-form text, which has implications for how educators and administrators communicate with them (Hampton & Keys, 2017). Educators and administrators should be mindful of these preferences when communicating with Gen Z students and consider adopting digital platforms, such as social media, to engage with them more effectively.

Although Gen Z students are known for their adeptness in communicating through mobile devices, they often encounter difficulties in face-to-face communication (Hernandez-de-Menendez et al., 2020) and tend to favour individual learning environments that allow them to focus and maintain a personalized pace (Seemiller & Grace, 2017). This inclination towards intrapersonal learning is distinct from Millennials' interpersonal approach, which emphasizes collaboration and teamwork (Seemiller & Grace, 2017). This perspective aligns with Lai and Hong's (2015) findings, which noted that university students born after 1992 had a significantly different preference for group work compared to students born between 1982-1992 and those born before 1982. Interestingly, Lai and Hong (2015) found no significant differences among the three cohorts in terms of their enjoyment of discussing ideas with fellow students. This suggests that, although Gen Z students may prefer intrapersonal learning, they are willing to engage in conversations with peers and can collaborate when necessary (Isaacs et al., 2020).

Gen Z students expect their educational experiences to be technology-driven and relevant to the real world. They value practical, active hands-on learning experiences that integrate technology and prepare them for the workforce. They are independent learners (Eckleberry-Hunt, Lick, & Hunt, 2018; Schwieger & Ladwig, 2018) and are generally more socially and politically engaged than previous generations, with a strong focus on social justice and activism (Seemiller & Grace, 2016). According to EAB research (2019), Gen Z and Millennial students may have become more self-sufficient learners, relying on technology to solve problems and find information. Its study shows that only 3 out of 10 respondents from both groups indicated that they would seek assistance from a teacher before attempting to resolve a problem on their own or using online resources. This shift towards independent learning and technology usage could have implications for the design and delivery of educational programmes aimed at engaging and supporting these generations. Additionally, Gen Z have an attention span of only 8 seconds on average (Glum, 2015) and often expect immediate feedback. Bíró's (2014) and Borys and Laskowski's (2013) studies have both highlighted that gamification can enhance Gen Z engagement in the learning process for individuals through the use of group-based motivation and feedback mechanisms.

Overall, Gen Z individuals are characterized by their entrepreneurial problem-solving skills, rapid information access, and adaptability to change. They value living in the present, with a focus on immediate satisfaction and superficial virtual relationships (Hernandez-de-Menendez et al., 2020). They seek opportunities for personal and professional growth and expect their educational experiences to prepare them for success in the workforce; however, Gen Z may not have great critical thinking or information literacy skills (Bashri & Rafeeq, 2020; Marr, 2022; Seibert, 2020). Institutions should focus on creating learning environments that foster the development of digital literacy and technological skills, along with providing opportunities for students to apply their learning in real-world contexts.

### 2.2.5. Summary of the Generational Characteristics

The literature clearly indicates that there are significant differences in the teaching preferences, learning styles, technology usage, and communication methods among the four generations discussed above. The understanding of generational differences is particularly crucial for the effective adoption of GenAI technologies in higher education. As mentioned by Linnes and Metcalf (2017), it is important to understand the characteristics of Gen Z to incorporate their needs and preferences in educational settings, and with the increasing use of AI technologies in higher education, it is important for educators and policymakers to consider the unique features of each generation to provide effective support and guidance. For instance, Gen Z learners are known to be hyperconnected and facile with computers and the internet, which can be leveraged to develop new GenAI API technologies that align with their learning preferences. Meanwhile, Gen X and Gen Y teachers may have different comfort levels and experiences with these technologies, requiring different forms of training and support to effectively incorporate GenAI technologies in their teaching practices. By understanding these generational differences and adapting to them, higher education institutions can successfully integrate GenAI technologies to improve the quality of education and meet the evolving needs of each generation.

Table 1: An overview of distinct generational traits and characteristics

| CHARACTERISTIC | BABY-BOOMER ~ 1946 - 1960 | GEN X ~ 1960 - 1980 | GEN Y/ MILLENIALS ~ 1980 - 1995 | GEN Z ~ 1995 - 2012 |
|---|---|---|---|---|
| TEACHING PREFERENCE | Traditional lecture format | Combination of traditional and technology-based methods | Interactive, self-paced, technology-based methods | Hybrid (blended) learning, technology-focused |
| LEARNING STYLE | Teacher-centered, note-taking, memorization | Collaborative, project-based, real-world application | Collaborative and networked, technology-based | Learn through images/videos/audio instead of text<br>Experiential active Learning |
| TECHNOLOGY | Early information | Uses with confidence | Part of everyday life, intuitive | Digital-first |

|  | technology (IT) adaptors | Digital immigrants | Digital Natives | Technoholics |
|---|---|---|---|---|
| **SOCIAL MEDIA** | N/A | Some use for personal communication | High use for personal and professional communication | Integrated into daily life |
| **FEEDBACK** | Once per year, during the annual review | Weekly/daily | On demand | Consistent, immediate and frequent |
| **COMMUNICATIONS APPROACHES** | Telephone | Email and text messages | Text or social media | Hand-held communication devices |
| **COMMUNICATION PREFERENCE** | Face to face ideally, but telephone or email if required | Text messaging or email | Online and mobile texting | Facetime |
| **KNOWLEDGE SHARING** | Willingly, voluntarily | Based on mutuality and cooperation | Only in cases of self-interest or if forced | On virtual level, easily and rapidly, no stake, publicly |
| **VALUES** | Patience, soft skills, respect for traditions, EQ, hard work | Hard work, openness, respect for diversity, curiosity, practicality | Flexibility, mobility, broad but superficial knowledge, success orientation, creativity, freedom of information takes priority | Live for the present, rapid reaction to everything, initiator, brave, rapid information access and content search |
| **ATTITUDE TOWARDS CAREER** | Organisational – careers are defined by employers | Early "portfolio" careers – loyal to profession, not necessarily to employer | Digital entrepreneurs – work "with" organisations and not "for" organisations | Career multitaskers – will move seamlessly between organisations and "pop up" businesses |
| **AIM AND ASPIRATION** | Solid existence Job security | Multi-environ-ment, secure position Work life balance | Freedom and flexibility | Live for the Present |
| **RELATIONSHIP** | First and foremost personal | Personal and virtual networks | Principally virtual, network | Virtual and superficial |
| **VIEW** | Communal, unified thinking | Self-centered and medium-term | Egotistical, short-term | No sense of commitment, be happy with what you have and live for the present |
| **PROBLEM SOLVING** | Horizontal | Independent | Collaborative | Entrepreneurial |
| **TEAMWORK** | Unknown | Natural environment (multinational companies) | On a virtual level (only if forced) | Virtual and rapid |
| **CHANGE MANAGEMENT** | Change = caution | Change = opportunity | Change = improvement | Change = expected |
| **TRAINING** | Preferred in moderation | Required as necessary | Continuous and expected | Ongoing and essential |
| **BEHAVIOR** | Challenge the rules | Change the rules | Create the rules | Customize the rules |

Extracted and analysed from the following literature (Bencsik et al., 2016; Bíró, 2014; Borys & Laskowski, 2013; EAB, 2019; Eckleberry-Hunt, Lick, & Hunt, 2018; Glum, 2015; Granitz, Kohli, & Lancellotti, 2017; Hampton & Keys, 2017; Hernandez-de-Menendez et al., 2020; Issac et al., 2020; Linnes & Metcalf, 2017; Mosca, Curtis, & Savoth, 2019; Oblinger & Oblinger, 2005; Puiu, 2017; Schwieger & Ladwig, 2018; Shamma, 2011; Shatto & Erwin, 2016; Turner, 2015; Wiedmer, 2015; Zemke et al., 2000)

## 3. Methodology

This study employed an online survey method to investigate the current experiences, perceptions, knowledge, and concerns of Gen Z students and Gen X and Millennial teachers regarding the use of GenAI in higher education. The survey consisted of both open and closed questions, aiming to capture a comprehensive understanding of the participants' views.

A convenience sampling technique was used to recruit participants for the study. Bulk emails were sent to potential participants, inviting them to participate in the online survey. The informed consent form was provided on the online platform before the participants could access the survey questions, ensuring that they were aware of the study's objectives and their rights as participants. The collected data were analysed using a two-fold approach: first, descriptive analysis was utilized to examine the quantitative data from closed questions, providing insights into the participants' experiences, perceptions, knowledge, and concerns about GenAI in higher education. Next, thematic coding analysis was utilised to code, categorise, and make sense of the qualitative data obtained from the open-ended questions in the survey. This method allowed for the identification of recurring themes and patterns, offering a deeper understanding of the participants' views on the use of GenAI in higher education. The open-ended questions were:

- How will you use generative AI technologies like ChatGPT in your teaching and learning?
- How concerned are you about the rapid adoption and widespread use of AI technologies in various industries and aspects of society?
- Are there any other comments you would like to share?

### 3.1. Sample Size and Data Analysis

The demographic information of the 583 participants, of which there were 399 students and 184 teachers, are shown in Table 2. As previously mentioned, the majority of teachers generally fall within the birth year ranges of Gen X or Gen Y, and students typically belong to the Gen Z age group. Consequently, the survey did not include a specific question regarding the age of teachers, as the participants' generational affiliations were already assumed based on their roles as teachers or students. T-test analyses were conducted to identify any significant differences between students' and teachers' survey responses. Aside from Item 1 (on frequency of GenAI usage), the remaining 21 survey items allowed for a sixth Not Sure option. As this fall outside of the Strongly Disagree – Strongly Agree 5-point Likert scale, instances of Not Sure were treated as missing values during t-test analyses, then later revisited to see whether the proportion of uncertainty towards each survey item was significantly different between students and teachers (i.e., t-test analyses were run after recoding responses falling within the SD – SA scale as 0, and Not Sure coded as 1).

Answers to the three open-ended survey questions were analysed by two independent coders using inductive thematic coding. To establish the initial coding scheme as well as intercoder agreement, two coders independently coded a random selection of 50 participant responses and then later met to discuss disagreements and finalise the coding protocol. Percentage agreement is an acceptable alternative that reflects reliability (Feng, 2015), particularly as other reliability coefficients (e.g., Cohen's kappa) are difficult to calculate when there are different numbers of codes across different categories (Cheung & Tai, 2021). The intercoder agreement percentages were calculated as 72% for the coding conducted to identify participants' willingness and intentions to use GenAI technologies, and 77% for that done to identify participant's concerns regarding GenAI usage in higher education.

Table 2. Demographics of participants

|  | Students ($n$ = 399) | Teachers ($n$ = 184) | Total ($n$ = 583) |
|---|---|---|---|
| **Gender** |  |  |  |
| Male | 201 | 107 | 308 |
| Female | 198 | 77 | 275 |
| **Age** |  |  |  |
| Mean ± SD | 22.1 ± 2.59 | - | - |
| Median | 22 | - | - |
| Min-max | 17-28 | - | - |
| **Location of study/work** |  |  |  |
| Hong Kong | 349 | 157 | 506 |
| Mainland China | 33 | 2 | 35 |
| United Kingdom/Ireland | 8 | 8 | 16 |
| Australia | 2 | 3 | 5 |
| North America | 3 | 1 | 4 |
| East Asia | 2 | 0 | 2 |
| Developing | 0 | 12 | 12 |
| Not specified | 2 | 1 | 3 |
| **Discipline** |  |  |  |
| Architecture | 26 | 11 | 37 |
| Arts | 44 | 38 | 82 |
| Business | 55 | 6 | 61 |
| Dentistry | 1 | 0 | 1 |
| Education | 24 | 36 | 60 |
| Engineering | 123 | 16 | 139 |
| Law | 8 | 3 | 11 |
| Medicine | 7 | 17 | 24 |
| Science | 55 | 16 | 71 |
| Social Sciences | 22 | 16 | 38 |
| Not specified/Other | 34 | 25 | 59 |
| **Level of study** |  |  |  |
| Undergraduate (total) | 244 | - | - |
| Freshman | 81 | - | - |
| Sophomore | 51 | - | - |
| Junior | 48 | - | - |
| Senior | 58 | - | - |
| Fifth-sixth year | 6 | - | - |
| Taught postgraduate | 111 | - | - |
| Research postgraduate | 44 | - | - |
| **Level of work** |  |  |  |
| Professor (Professor, Associate, Assistant) | - | 65 | - |
| Lecturer (Assistant, Lecturer, Senior, Principal) | - | 73 | - |
| Post-doctorate | - | 6 | - |
| Teaching/research assistant | - | 16 | - |
| Other | - | 24 | - |

## 4. Findings

### 4.1. Quantitative results

Table 3 presents the results of the student's *t*-test comparing the item scores between Gen Z students and Gen X and Y teachers. In cases where Levene's test was significant ($p < .05$, signifying unequal variances), the adjusted Welch's *t* is reported. Overall, while no significant group differences were found, both students and teachers tended to agree with statements regarding the need for higher education institutions to establish a plan addressing GenAI use, the importance for students to be able to utilise GenAI effectively for their future careers, and several items regarding their awareness of GenAI's limitations and risks.

For items in which significant differences were found, students overall reported a greater frequency ($M_S$=2.27, $SD_S$=1.65; with a score of 1 corresponding to "Never" and 5 corresponding to "Always") of GenAI technologies usage, including ChatGPT, compared to teachers ($M_T$=2.03, $SD_T$=1.11; $t(581)$, $p$=.023). It could be that the younger members of Generation Z are more open and accustomed to trying and adopting new and upcoming technologies, especially with the exponential rise in the popularity of ChatGPT in particular.

Likewise, a more open-minded attitude could also help account for students' greater level of agreement ($M_S$=3.90, $SD_S$=.82), compared to that of their teachers ($M_T$=3.58, $SD_T$=1.11), that integrating GenAI technologies into higher education would positively impact teaching and learning in the future ($t(234)$=3.36, $p<.001$). Combined with the aforementioned higher frequencies of usage, which in itself could also indicate greater familiarity with how the technology can work and be used, students were more likely to believe that GenAI technologies could help them save time ($M_S$=4.16, $SD_S$=.83; $M_T$=3.90, $SD_T$=.88; $t(555)$=3.38, $p<.001$) and become better writers ($M_S$=3.29, $SD_S$=1.16; $M_T$=3.06, $SD_T$=1.23; $t(545)$=2.07; $p$=.039) compared to what teachers believed about the utility of GenAI for students. Furthermore, students were more likely to see GenAI technologies as a useful and good tool for student support services, given the anonymity it provides ($M_S$=3.73, $SD_S$=1.66; $M_T$=3.53, $SD_T$=1.11; $t(536)$=2.07; $p$=.039), with their confidence in what GenAI is capable of also resulting in a higher level of agreement, compared to teachers, that such technologies were unlikely to be affected by harmful input that will distort or manipulate its outputs ($M_S$=2.97, $SD_S$=1.22; $M_T$=2.57, $SD_T$=1.28; $t(510)$=3.35; $p<.001$).

On the other hand, results suggested that teachers may be more skeptical of the capabilities of GenAI, as well as concerned about the risks and dangers it poses to students' learning, growth, and academic achievements. Teachers showed a greater level of caution regarding GenAI outputs, scoring higher in their agreement with the need to fact-check and validate information produced by GenAI technologies ($M_S$=4.35, $SD_S$=.81; $M_T$=4.60, $SD_T$=.65; $t(402)$=-3.95; $p<.001$); they also tended to score higher than students in their agreement with GenAI having the potential of generating factually inaccurate outputs

($M_S$=4.08 $SD_S$=.85; $M_T$=4.27, $SD_T$=.78; $t(554)$=-2.54; $p$=.011) and outputs that exhibit biases and unfairness ($M_S$=3.91, $SD_S$=.93; $M_T$=4.20, $SD_T$=.78; $t(545)$=-3.52; $p$<.001), as well as with GenAI impeding students' opportunities to interact and socialise with others ($M_S$=3.09, $SD_S$=1.19; $M_T$=3.43, $SD_T$=1.14; $t(542)$=-3.16; $p$=.002).

Finally, teachers further expressed a greater level of concern for the possibility of some students using GenAI technologies to get ahead in their assignments ($M_S$=3.58, $SD_S$=1.14; $M_T$=3.83, $SD_T$=1.06; $t(366)$=-2.51; $p$=.013), and were more likely to believe that students would become over-reliant on GenAI ($M_S$=2.87, $SD_S$=1.14; $M_T$=4.12, $SD_T$=.89; $t(425)$=-1.25; $p$<.001) compared to what students thought about themselves.

Table 3: Descriptive data and *t*-test analysis results

| Item | Students | | Teachers | | MD | t | (df) | P |
|---|---|---|---|---|---|---|---|---|
| | n | M(SD) | n | M(SD) | | | | |
| 1. I have used generative AI technologies (GenAI) like ChatGPT. | 399 | 2.27 (1.65) | 184 | 2.03 (1.11) | .23 | 2.28 | 581 | .023* |
| ***Perceptions*** | | | | | | | | |
| 2. The integration of GenAI like ChatGPT in higher education will have a positive impact on teaching and learning in the long run.[a] | 383 | 3.90 (.82) | 160 | 3.58 (1.11) | .33 | 3.36 | 234 | <.001* |
| 3. Higher education institutions should have a plan in place for managing the potential risks associated with using GenAI like ChatGPT in teaching and learning. | 393 | 4.47 (.83) | 182 | 4.53 (.86) | -.06 | -.79 | 573 | .432 |
| 4. Students must learn how to use GenAI well for their career. | 387 | 3.97 (.98) | 171 | 3.96 (.98) | .00 | .02 | 556 | .987 |
| 5. I believe using GenAI like ChatGPT to write essays or generate answers to questions can lead to originality and creativity in my/students' work. | 379 | 3.09 (1.20) | 161 | 2.98 (1.30) | .11 | .91 | 538 | .362 |
| 6. I believe GenAI like ChatGPT can improve my/students' digital competence. | 377 | 3.66 (.96) | 172 | 3.68 (.98) | -.03 | -.28 | 547 | .778 |
| 7. I believe GenAI like ChatGPT can help me/students save time. | 386 | 4.16 (.83) | 171 | 3.90 (.88) | .26 | 3.38 | 555 | <.001* |
| 8. I think GenAI like ChatGPT can help me/students become a better writer. | 379 | 3.29 (1.16) | 168 | 3.06 (1.23) | .23 | 2.07 | 545 | .039* |
| 9. I/Students will not feel judged by GenAI like ChatGPT, so I/they feel comfortable with it.[a] | 377 | 3.54 (1.05) | 159 | 3.68 (.92) | -.14 | -1.56 | 337 | .121 |
| 10. I think AI technologies like ChatGPT is a great tool for student support services due to anonymity. | 372 | 3.73 (1.01) | 166 | 3.53 (1.11) | .20 | 2.07 | 536 | .039* |
| ***Knowledge*** | | | | | | | | |
| 11. I understand GenAI like ChatGPT have limitations in their ability to handle complex tasks. | 391 | 4.13 (.84) | 168 | 4.15 (.78) | -.03 | -.39 | 557 | .699 |
| 12. … can generate output that is factually inaccurate. | 382 | 4.08 (.85) | 174 | 4.27 (.78) | -.19 | -2.54 | 554 | .011* |
| 13. … can generate output that is out of context or inappropriate. | 385 | 4.01 (.85) | 174 | 4.14 (.82) | -.14 | -1.80 | 557 | .072 |
| 14. … can exhibit biases and unfairness in their output. | 380 | 3.91 (.93) | 167 | 4.20 (.78) | -.29 | -3.52 | 545 | <.001* |
| 15. … may rely too heavily on statistics, which can limit their usefulness in certain contexts. | 381 | 3.96 (.91) | 162 | 3.86 (.94) | .09 | 1.12 | 541 | .265 |
| 16. … have limited emotional intelligence and empathy, which can lead to output that is insensitive or inappropriate. | 380 | 3.87 (.98) | 163 | 3.98 (1.01) | -.11 | -1.16 | 541 | .247 |
| 17. … cannot be affected by harmful input, and will not cause the output to be distorted or manipulated. | 361 | 2.97 (1.22) | 151 | 2.57 (1.28) | .40 | 3.35 | 510 | <.001* |
| 18. I think it's important to fact-check and validate information generated by GenAI like ChatGPT, before using it for assignments.[a] | 389 | 4.35 (.81) | 173 | 4.60 (.65) | -.25 | -3.95 | 402 | <.001* |
| *Concerns* | | | | | | | | |
| 19. I am concerned that some students may use GenAI like ChatGPT to get ahead in their assignments.[a] | 389 | 3.58 (1.14) | 177 | 3.83 (1.06) | -.25 | -2.51 | 366 | .013* |
| 20. Using GenAI like ChatGPT to complete assignments undermines the value of a university education. | 382 | 3.18 (1.16) | 174 | 3.40 (1.20) | -.21 | -1.95 | 555 | .052 |
| 21. GenAI like ChatGPT will limit my/students' opportunities to interact with others and socialise when completing coursework. | 383 | 3.09 (1.19) | 168 | 3.43 (1.14) | -.34 | -3.16 | 542 | .002* |
| 22. I/Students can become over-reliant on GenAI.[a] | 368 | 2.87 (1.14) | 174 | 4.12 (.89) | -1.25 | -14.00 | 425 | <.001* |

*Significant difference $p < .05$

Note[a]: Levene's test for the Student's *t*-test was significant ($p < .05$), the adjusted Welch's *t* is thus reported for this item.

### 4.1.1. Uncertainty Among Students and Teachers

With the novelty of GenAI technologies like ChatGPT for use both in academia and in general, the *Not Sure* response option was included in our survey to account for any uncertainty among participants, such as for those who felt that they still lacked adequate information to form a stronger opinion (Krosnick & Presser, 2015). Table 4 presents the percentages of participants' *Not Sure* responses and t-test comparisons of such between students and teachers.

Significant differences were found between students and teachers in a number of Perception and Knowledge items, where in comparison with the former, a greater proportion of the latter responded with uncertainty towards the benefits of GenAI integration in higher education for the future of teaching and learning ($t(242)=-3.38$, $p<.001$); the usage of GenAI for increasing students' originality and creativity ($t(256)=-2.90$; $p=.004$) or helping students save time ($t(242)=-2.35$; $p=.020$); and whether students do not feel judged by GenAI technologies and thus feel more comfortable with using it ($t(245)=-3.39$; $p<.001$).

Furthermore, a significantly different and greater proportion of teachers expressed uncertainty regarding the capabilities and limitations of GenAI technologies, including the extent to which it can handle complex tasks ($t(221)=-3.18$; $p=.002$), exhibit biases and unfairness ($t(279)=-1.98$; $p=.049$), over-rely on statistics ($t(247)=-3.06$; $p=.002$), demonstrate or exercise emotional intelligence and empathy ($t(250)=-2.89$; $p=.004$), and the extent to which it can*not* be affected by harmful input that distorts or manipulates its outputs ($t(283)=-2.71$; $p=.007$).

Table 4: Frequencies and *t*-test comparisons of "Not Sure" responses

|  | Students | | Teachers | | | | | |
|---|---|---|---|---|---|---|---|---|
| **Item** | n | % "Not sure" | n | % "Not sure" | MD | t | (*df*) | *p* |
| *Perceptions* | | | | | | | | |
| Item 2[a] | 399 | 4.01 | 184 | 13.04 | -.09 | -3.38 | 242 | <.001* |
| Item 3 | 399 | 1.50 | 184 | 1.09 | .00 | .40 | 581 | .688 |
| Item 4[a] | 399 | 3.01 | 184 | 7.07 | -.04 | -1.95 | 260 | .052 |
| Item 5[a] | 398 | 4.77 | 184 | 12.50 | -.08 | -2.90 | 256 | .004* |
| Item 6 | 396 | 4.80 | 184 | 6.52 | -.02 | -.86 | 578 | .391 |
| Item 7[a] | 395 | 2.28 | 184 | 7.07 | -.05 | -2.35 | 242 | .020* |
| Item 8[a] | 397 | 4.53 | 184 | 8.70 | -.04 | -1.79 | 279 | .175 |
| Item 9[a] | 394 | 4.31 | 184 | 13.59 | -.09 | -3.39 | 245 | <.001* |
| Item 10[a] | 397 | 6.30 | 184 | 9.78 | -.03 | -1.39 | 300 | .166 |
| *Knowledge* | | | | | | | | |
| Item 11[a] | 398 | 1.76 | 184 | 8.70 | -.07 | -3.18 | 221 | .002* |
| Item 12 | 398 | 4.02 | 184 | 5.43 | -.01 | -.77 | 580 | .443 |
| Item 13[a] | 396 | 2.78 | 184 | 5.43 | -.03 | -1.42 | 275 | .156 |
| Item 14[a] | 398 | 4.52 | 184 | 9.24 | -.05 | -1.98 | 279 | .049* |
| Item 15[a] | 397 | 4.03 | 184 | 11.96 | -.08 | -3.06 | 247 | .002* |
| Item 16[a] | 396 | 4.04 | 184 | 11.41 | -.07 | -2.89 | 250 | .004* |
| Item 17[a] | 398 | 9.30 | 184 | 17.93 | -.09 | -2.71 | 283 | .007* |
| Item 18[a] | 398 | 2.26 | 184 | 5.98 | -.04 | -1.95 | 252 | .052 |
| *Concerns* | | | | | | | | |
| Item 19 | 396 | 4.02 | 184 | 3.80 | .00 | .12 | 580 | .901 |
| Item 20[a] | 397 | 3.53 | 184 | 5.43 | -.02 | -1.00 | 299 | .320 |
| Item 21[a] | 396 | 5.05 | 184 | 8.70 | -.04 | -1.55 | 289 | .123 |
| Item 22 | 396 | 7.07 | 184 | 5.43 | .02 | .74 | 578 | .460 |

*Significant difference $p < .05$

Note[a]: Levene's test for the Student's *t*-test was significant ($p < .05$), the adjusted Welch's *t* is thus reported for this item.

## 4.2. Qualitative results

### 4.2.1. Willingness and Intentions

In general, there were many similarities in the ways that both Gen Z students and Gen X and Y teachers intended to utilise GenAI technologies for and within higher education. Among the most prominent codes and themes were the intentions of both groups to use such technologies for acquiring, compiling, and consolidating information, including for brainstorming and gaining inspiration, condensing and summarising complex ideas, as well as utilising GenAI to help construct write literature reviews for research papers and assisting with data analysis. Other common intentions of use included language learning, using it for simple and repetitive tasks such as administrative work, and increasing one's productivity and efficiency overall. Writing support was also a common theme, both in terms of technical writing (e.g., essay structure and organisation, grammatical checks, proofreading) and writing support for specific assignments and in general (e.g., improving arguments, improving one's own writing skills).

Additionally, both groups had intentions to use GenAI technologies as a form of personalised and immediate support and feedback for teaching and learning. For example, some students said they would use it for themselves to expand their existing knowledge, find learning resources and materials for practice (e.g., running mock interviews with GenAI), or use the technology to generate feedback about their own writing and assignments before submission. Likewise, while teachers might use it to develop teaching activities, plans, and materials, some were also happy to allow and even encourage students to use GenAI for such purposes. In fact, a number of teachers intended to specifically teach their students about using technologies for academic work in a more effective and responsible manner, such as emphasising its current capabilities and limitations, with others going further, intending to have their students directly explore, critique, and reflect on the outputs produced by GenAI technologies in order to directly foster critical thinking and evaluative skills.

However, some participants did believe that GenAI usage would hinder or even defeat the purpose of teaching and learning in certain areas, such as in the field of language and writing itself. Interestingly, among the several participants who said they would not, or had no intention of using GenAI at the moment, students cited reasons such as complying with current university regulations or in one case, due to a strong dislike for such technologies as they posed a threat to creativity and humanity. On the other hand, for teachers who had no current intentions to make use of GenAI, they were more so unimpressed – some believed that GenAI capabilities were currently overestimated and lacked "intelligence", given that it does not generate new content as much as it compiles and repackages existing information from its database.

### 4.2.2. Concerns

In addition to those addressed in the quantitative findings, Gen Z students and Gen X and Y teachers expressed a number of similar concerns regarding the adoption and use of GenAI technologies in higher education. These include over any unethical, dishonest, and irresponsible uses of GenAI including cheating and plagiarism, as well as regarding the potential of GenAI producing material that is low quality or includes misinformation and biases, particularly if it perpetuates or exacerbates social injustices and inequality. Many also expressed worries about the larger impact of such technologies on the job market and society, with both students and teachers anxious or distressed about job losses or the potential of "humans being replaced" in the future, as well as the undermining of academic degrees and integrity, privacy and transparency concerns, and any threats GenAI may pose to society and human values should it come to develop its own, misaligned set of values.

Some teachers believed that students may not "have enough knowledge to identify the reliability of information generated by GenAI technologies", particularly undergraduate students in their earlier years that may not have had adequate practice and exposure to evaluating information for accuracy and validity. They expressed their existing lack of confidence in students' current critical thinking and evaluative abilities, and when combined with the belief that GenAI use and dependence would "hamper the development of essential skills needed" in the workplace and society, some had a bleak outlook on how these technologies might restrict and obstruct learning:

> *"I believe over-reliance on AIs in learning will adversely affect students' learning as [they] may be deprived of the actual thinking process which is the most important part of learning."*

In a similar vein, teachers in particular raised concerns about current assessment practices and criteria, the need to re-consider them in light of new GenAI tools available to students. While GenAI technologies may be good at helping students "regurgitate information", learning is much more than that, incorporating the development of analytical and speaking skills, creativity, reflexivity, applying knowledge and concepts to personal and new experiences, and so on. Students could continue to be tested on their basic knowledge, but other assessment tasks and their designs will need to consider the different GenAI technologies that currently do exist, and later will exist. This is especially necessary given the shift away from traditional examinations taking place in supervised, in-person settings ever since the Covid-19 pandemic. What should be and needs to be assessed will need to be carefully considered by teachers and curriculum designers in the near future; though this could include, for example, assessing students on their ability to use GenAI to write and improve academic papers, ill-designed and traditional tasks may not best reflect students' actual effort and work:

> *"It could be a problem for teachers to detect this usage, and it could allow students to get good grades without actually possessing the skills that those grades reflect."*

At the same time, it is still important to recognise current limitations of GenAI, not only in terms of how they may be used unfairly by students, but also how easily false detections of unfair GenAI usage may occur. One teacher shared their use of multiple programmes meant to detect whether a piece of writing was generated by AI technology; after running flagged pieces though other similar programmes and speaking to students individually, they concluded that it was highly unlikely that students whose writing were flagged had intended to cheat or use GenAI to do their assignments for them. Some had in fact used it to improve their grammar or translate their original writing, completed in their native language, into

English, while others were seemingly flagged for no reason. Along with the fact that all the flagged pieces of writing were already poor quality, it seems that current programmes meant to detect writing generated by AI is highly unreliable and should also be taken into consideration.

Still, many do accept that GenAI is part of the evolution of technology and its trends, comparing current apprehensions and concerns over it to historical developments and reactions to technology like radios, televisions, and even the Internet. A reoccurring theme for both groups of respondents was participants' emphasis on the importance of being informed about GenAI technologies and both their potentials and limitations, as well as the need for continued discussion about how to handle and oversee such technologies in education. Many agreed that guidelines and protocols for fair and responsible use are urgently needed, particularly with multiple comments from participants regarding how the "impact of GenAI depends on the intentions of its users". Moreover, though resistance and reluctance still existed in both groups, students and teachers did seem to agree that banning or prohibiting GenAI would be more harmful than good, and that "the future is to 'collaborate' with [these technologies], not to ban them", otherwise it would be "detrimental to our students' future as others will certainly be using this tool to refine their writing, thinking, and so forth."

Likewise, some participants did recognise that there may be a need for both students and teachers to reskill in order to keep up with the future of GenAI technologies, such as in terms of digital literacy. Many did not want to fall behind, not only in keeping up with new technologies but also behind others – individuals, institutions, and even countries – in adopting and making use of such powerful tools; several respondents also pointed out that fair access and digital poverty are also necessary talking points when establishing plans and protocols for GenAI adoption in education. Moreover, some teachers had further concerns over how quickly and rapidly the adoption of GenAI was taking place without any proper policies or guidelines for its usage in place; without this, and even dedicated teams to help educators keep up with developments, teachers already have too much on their plate; as such, some participants called for "a clear policy ... to help our students keep with the trend instead of simply banning it". Another added,

> "*I think there needs to be a clear strategy in place ASAP, as many core subjects lend themselves to the use of AI. I think we need to teach the students how to use AI tech well and responsibly.*"

Finally, there was still excitement about the potential of GenAI despite the aforementioned concerns, with some believing that uncertainties and anxieties would lessen over time as we become more familiar and accustomed to these technologies. Several respondents believed that humans would always be needed and be part of the loop in managing GenAI development and uses, and for teachers themselves, some did believe that their input in teaching and providing feedback would always still be needed.

> *"I am curious to learn more about it and to participate in discussions on how to maximize its use without compromising academic integrity and students' learning."*

## 5. Discussion

This study explored the experiences, perceptions, knowledge, concerns, and intentions of Gen Z students in comparison with their Gen X and Gen Y teachers regarding the use of generative AI in higher education. The findings offered valuable insights into the participants' attitudes towards GenAI, and the potential implications of integrating this technology into educational settings.

### 5.1. GenAI Perceptions across the Generations

Gen Z students, having grown up with technology and the internet, are more likely to embrace new technological advancements such as GenAI. Our findings support this notion, as Gen Z participants demonstrated optimism towards the potential benefits of GenAI in higher education, including enhancing productivity, efficiency, and personalized learning. Furthermore, Gen Z students expressed intentions to use GenAI for various educational purposes, such as acquiring and consolidating information, language learning, and writing support, aligning with previous research suggesting that they value technology as a means to enhance their learning experiences (Hernandez-de-Menendez, Escobar Diaz & Morales-Menendez, 2020; Seemiller & Grace, 2016).

According to Marshall and Wolanskyj-Spinner (2020), Gen Z students are "active problem solvers, independent learners," which makes them particularly suitable to adopt GenAI. GenAI can also fulfill the expectation of Gen Z students for constant quick personalized feedback (Bíró, 2014; Borys & Laskowski, 2013). This trend towards independent learning may be further facilitated by the development of GenAI technologies, such as chatbots and intelligent tutoring systems, which can provide personalized and immediate feedback to students, encouraging them to rely less on teachers for support. Moreover, GenAI systems can adapt to a student's learning style and pace, providing them with targeted feedback and resources. This can be particularly beneficial for Gen Z students, who are used to having access to personalized and on-demand services through technology with zero tolerance for delays.

On the other hand, Gen X and Gen Y teachers, who have experienced the transition from traditional to technology-based educational settings, demonstrated a more cautious approach to GenAI adoption. They acknowledged the potential benefits of GenAI but showed greater uncertainty and concerns about ethical and pedagogical implications, emphasizing the need for proper guidelines and policies to ensure responsible use of the technology. This is consistent with previous research suggesting that Gen X (a more self-reliant generation) and Gen Y individuals tend to be more skeptical of new technologies, placing a greater emphasis on the potential risks and challenges associated with their adoption. The quantitative data in table 3 supports this, showing that teachers generally have more concerns about adopting GenAI in teaching and learning than students. For instance, teachers perceive a lower positive impact of GenAI (*Item 2; $M_T$ =3.58 vs. $M_S$ =3.90*), are more concerned about students using GenAI to get ahead (*Item 19; $M_T$ =3.83 vs. $M_S$ =3.58*), worry more about GenAI limiting students' social interactions (*Item 21; $M_T$ =3.43 vs. $M_S$ =3.09*), and are more concerned about students becoming over-reliant on GenAI (*Item 22; $M_T$ =4.12 vs. $M_S$ =2.87*).

**5.2. Concerns and Considerations for the Future**

The concerns raised by both the Gen Z students and the Gen X and Gen Y teachers about plagiarism, misinformation, and biased content also echo existing literature on the ethical challenges posed by AI in education (Gillissen et al., 2022; Jha et al., 2022). They recognize GenAI's limitations, such as generating inaccurate output (*Item 12; $M_T$ =4.27; $M_S$= 4.08*) and out-of-context or inappropriate content (*Item 13; $M_T$ =4.14; $M_S$=4.01*). Both groups understand that GenAI can exhibit biases (*Item 14; $M_T$ = 4.20; $M_S$=3.91*) and may over-rely on statistics (*Item 15; $M_T$ = 3.86; $M_S$= 3.96*). They also acknowledge GenAI's limited emotional intelligence and empathy (*Item 16; $M_T$ =3.98; $M_S$=3.87*). Despite some differences, both students and teachers share concerns about GenAI's limitations in producing accurate, high-quality content. Given that literature has found Gen Z to be more likely to trust information found online due their lack of critical thinking skills (Seemiller & Grace, 2016), it is important to ensure students are able to recognise instances in which information may be disingenuous or inaccurate and when fact-checking and validation is necessary. Teachers from Gen X and Gen Y, who value curiosity, creativity, and soft skills, are further concerned that students' overreliance on GenAI may hinder their skills development, including their capacity to think and evaluate information for themselves. Taken together, the participants' emphasis on the importance of ethical and responsible use of GenAI and the need for clear guidelines and protocols highlights the critical role of educators and policymakers in ensuring that GenAI integration maintains academic integrity and promotes equitable learning experiences. At the same time, educational institutions must also look to foster critical thinking and digital literacy skills in students, ensuring that they are able to evaluate the credibility of information and make responsible use of GenAI technologies.

Finally, our findings indicate that Gen Z students are generally more socially and politically engaged than previous generations, with a strong focus on social justice and activism (Seemiller & Grace, 2016). This underscores the importance of addressing the issues related to governance, particularly in terms of perpetuating or exacerbating social injustices and inequality. To this end, it is crucial for educators and policymakers to develop strategies for integrating GenAI technologies in a manner that promotes equitable learning experiences and social responsibility.

It is important to note that GenAI technologies are not a replacement for human teachers (Chan & Tsi, 2023). While AI systems can provide personalized feedback and resources, they may not be able to provide the same level of emotional support and social interaction as human teachers. Moreover, the development and use of AI systems should be done in a way that supports, rather than replaces, human teachers and their roles in education. This view is supported by both Gen Z students and Gen X and Gen Y teachers.

Although the majority of participants seem to be accepting and curious about the potential role of GenAI in the future of education and society, there is still uncertainty and resistance towards GenAI. As GenAI technologies continue to evolve, it is essential for individuals to stay informed about their potential benefits, risks, and ways to use them effectively, and for the development of guidelines, policies, regulations, as well as strategies, to ensure responsible and ethical use.

### 5.3. Limitations
This study has several limitations that should be acknowledged. First, the assumption that Gen X and Gen Y participants are teachers may not be entirely accurate; future studies should obtain participants' ages for more precise generational categorization. Second, the majority of the students and teachers in the study are from Hong Kong, which may limit the generalizability of the findings to other cultural and educational contexts. Likewise, generational studies are often geographically dependent, and most of the literature in this area is based on Western populations, potentially limiting the applicability of these findings to other regions. Additionally, the sample size was relatively small, further limiting the generalizability of the results. Future research should consider larger, more diverse samples from multiple institutions and countries, and investigate the cultural nuances that may influence attitudes and intentions towards GenAI adoption in higher education. Lastly, the reliance on self-reported data in this study may be subject to social desirability bias, which could affect the accuracy of participants' responses.

**6. Conclusions**

The adoption of GenAI technologies among Gen X, Gen Y (Millennials), and Gen Z varies due to the unique generational experiences and levels of familiarity with technology among each group. As AI becomes increasingly pervasive in various aspects of life, including education, work, and social interaction, it is essential to understand how different generations might interact with and adopt these emerging technologies into their lives.

Gen X has experienced the transition from analog to digital technologies during their lifetime. As "digital immigrants," they have had to adapt to new technological advancements, including AI, but may not be as inherently familiar with these technologies as their younger counterparts. Nonetheless, their resilience and adaptability, coupled with professional experience, could enable them to recognize the potential benefits of AI in solving complex problems, enhancing productivity, and improving decision-making processes. Gen X may initially approach AI with some skepticism, but their pragmatic nature will likely lead to its eventual acceptance and adoption if it proves valuable in various contexts.

Millennials, or Gen Y, are considered "digital natives" as they grew up during the rapid expansion of the internet and digital technology. They are generally more comfortable with technology and are likely to be early adopters of AI-powered tools and applications. Millennials value flexibility, efficiency, and innovation, which aligns with the capabilities offered by AI technologies. They are poised to leverage AI in various aspects of their lives, including career development, personal growth, and social engagement. Moreover, as digital entrepreneurs, they could drive the creation of new AI-based products and services that cater to the unique needs and preferences of their generation.

Finally, as mentioned, Gen Z is the first generation to grow up with constant access to digital technology, social media, and the internet. As a result, they are considered "digital-first" and "technoholic," with an inherent affinity for AI technologies. Gen Z is likely to embrace AI in various areas from work to daily life, due to their potential to enhance their efficiency, connectivity, and access to information. Their strong inclination towards visual learning, rapid information access, and multitasking abilities make them well-suited for adopting AI technologies that cater to these preferences. Furthermore, their entrepreneurial problem-solving mindset and adaptability to change could lead to the development of innovative AI solutions that address pressing challenges faced by society.

This study highlighted the importance of technology in education because it has the potential to enhance the learning experience of students. It provided evidence that traditional teaching methods and technology should be used together because they complement each other's strengths (Chan & Hu, 2023). While technology can provide interactive and engaging experiences for students, other teaching methods such as experiential learning (Chan, 2023) can help develop critical thinking and problem-solving skills. Therefore, combining the two can result in a more effective learning experience. Although Gen Z students are tech natives and have grown up with technology, it does not necessarily mean they prefer a tech-only approach.

Overall, the findings of this study contribute to the growing body of literature on the use of AI in education and provide important insights into the attitudes and intentions of Gen Z students and Gen X and Gen Y teachers towards the adoption of GenAI in educational settings. As GenAI continues to evolve, further research is needed to examine the long-term impact of GenAI integration on teaching and learning outcomes and to develop evidence-based guidelines and policies that promote responsible use of this technology in higher education. Additionally, educational institutions must foster critical thinking and digital literacy skills in students, ensuring that they are able to evaluate the credibility of information and use GenAI technologies in a responsible and ethical manner.

**Declarations:**

Availability of data and material: The datasets used and/or analysed during the current study are available from the corresponding author on reasonable request

Funding: No funding has been received for this study

Acknowledgements: The author wishes to thank the students and teachers who participated the survey.